\documentclass[aps,twocolumn,showpacs,superscriptaddress,amsmath,prb]{revtex4}
\usepackage{graphicx}

\begin{document}
\date{\today}
\title{Phase rigidity and 
avoided level crossings in the complex energy plane
}
\author{Evgeny N. Bulgakov}
\affiliation{Max-Planck-Institut f\"ur Physik komplexer
Systeme, D-01187 Dresden, Germany }
\affiliation{Kirensky
Institute of Physics, 660036 Krasnoyarsk, Russia}
\author{Ingrid Rotter}
\affiliation{Max-Planck-Institut f\"ur Physik komplexer
Systeme, D-01187 Dresden, Germany }
\author{Almas F. Sadreev}
\affiliation{Max-Planck-Institut f\"ur Physik komplexer
Systeme, D-01187 Dresden, Germany }
\affiliation{Kirensky
Institute of Physics, 660036 Krasnoyarsk, Russia}
\affiliation{Department of Physics and Measurement,\\
Technology Link\"{o}ping  University,  S-581 83 Link\"{o}ping,
Sweden}

\begin{abstract}
We consider the effective Hamiltonian of an open quantum system,
its biorthogonal eigenfunctions $\phi_\lambda$ and define the value 
$r_\lambda = 
(\phi_\lambda|\phi_\lambda)/\langle\phi_\lambda|\phi_\lambda\rangle$  
that characterizes the phase rigidity of the eigenfunctions $\phi_\lambda$.
In the scenario with avoided level crossings, $r_\lambda$
varies between 1 and 0 due to the mutual influence of neighboring resonances.
The variation of $r_\lambda$ may be considered 
as an internal property of an {\it open} quantum system.   
In the literature, the phase rigidity $\rho$ of the scattering wave function 
$\Psi^E_C$ is considered. Since 
$\Psi^E_C$ can be represented in the interior of the  system 
by the $\phi_\lambda$, the phase rigidity $\rho$ of the 
$\Psi^E_C$ is  related to the $r_\lambda$ and therefore also to the 
mutual influence of neighboring resonances.
As a consequence, the reduction of the phase rigidity  $\rho$ 
to values smaller than 1 should be considered, at least partly, 
as an internal property of an open quantum system in 
the overlapping regime.
The relation to measurable values such as the transmission through a quantum
dot, follows from the fact that the transmission is, in any case,
resonant with respect to the effective Hamiltonian.
We illustrate the relation between phase rigidity $\rho$ and transmission
numerically for small open cavities. 

\end{abstract}
\pacs{03.65.Yz, 73.23.-b, 73.63.Kv, 05.60.Gg}

\maketitle

\section{Introduction}

The advances in the nanotechnology make it possible to produce small quantum
dots  with desired controllable properties. The analogy between such a
system and an atom has proved to be quite close 
\cite{heiblum}. Since however no natural restrictions for choosing 
the control parameters exist,
the quantum dots may show new properties that we cannot obtain from 
studies on atoms. For example, Fano resonances have been observed  
experimentally in quantum dots \cite{gores}, and the Fano parameters 
may be complex \cite{kob}. Theoretically, the Fano parameter expresses the
interference between the resonant part of the transmission and a smooth 
(direct) nonresonant part. It is  real when the resonance part is caused by
the existence of an isolated resonance state \cite{fano}
as it is the case usually in atoms. When  however
the resonant part itself results from the interference of, e.g., two
neighboring resonance states, the Fano parameter 
becomes complex \cite{marost4}. The same may
appear in a cavity due to the absorption at the walls \cite{stefan2}. 

Further, it has been stated \cite{brouwer2} that the line shape 
of a Fano resonance may be affected by some dephasing \cite{definition} 
that may be caused by  intrinsic sources (e.g., from electron-electron
interactions) as well as by extrinsic sources 
(e.g., radiation, magnetic impurities) \cite{brouwer6}.
For a quantitative study, the phase rigidity  \cite{brouwer1,brouwer3}
\begin{eqnarray}
\rho =\frac{\int dr ~\Psi(r)^2}{\int dr ~|\Psi(r)|^2}
= e^{2i\theta} \frac{\int dr ~(|{\rm Re}\tilde\Psi(r)|^2 - 
|{\rm Im}\tilde\Psi(r)|^2)}{
\int dr ~(|{\rm Re}\tilde\Psi(r)|^2 + |{\rm Im}\tilde\Psi(r)|^2)} 
\label{rig1}
\end{eqnarray}
has been introduced that characterizes the degree to which the wave function 
$\Psi$ is really complex (the phase $\theta$ arises from a transformation 
so that Re$\tilde\Psi$ and Im$\tilde\Psi$ are orthogonal).  
An experimental and theoretical study \cite{stefan2} showed that the
two different mechanisms, dephasing and dissipation, are equivalent in terms
of their effect onto the evolution of Fano resonance line shapes.

Another interesting observation in the experimental results on quantum dots
\cite{gores,brouwer2} is that, as a
function of the gate voltage that controls the transparencies of the point
contacts, the widths of the observed resonances behave non-monotonic. The 
conductance peaks start as narrow Breit-Wigner resonances when the 
quantum dot is
pinched off, then widen as the contacts are opened into resonances exhibiting
the Kondo effect \cite{brouwer2}. As the contacts are opened further, the
resonances become more narrow and have  the Fano form with some background
conductance. A similar result is obtained in a study on a tunable microwave
scattering device \cite{stefan1}.
The explanation given in Ref. \cite{brouwer2} is that diffraction
at the contacts to the quantum dots is strongest at intermediate point contact
transparencies, leading to large sticking probabilities. 

The non-monotonic 
increase of resonance widths as a function of the degree of opening the 
system is however a typical feature of open quantum systems \cite{rep}. 
It was found first in theoretical nuclear reaction studies 
by using the formalism of the effective non-hermitian Hamiltonian 
$H_{\rm eff}$ \cite{kleinw}, 
then in atoms \cite{marost} and in microwave cavities \cite{persson} 
where it is proven  experimentally \cite{perssonexp,stefan1,stefan2}.
Also the transmission through a double quantum dot is studied 
as a function of the coupling strength between dot and attached leads
\cite{rstrans}. In the theoretical studies, 
the non-monotonic behavior of the resonance widths is 
caused by the width bifurcation that may appear at the
avoided  crossings of resonance states in the complex energy plane.
Further studies of open quantum systems showed 
that the real and imaginary parts of the eigenfunctions of 
$H_{\rm eff}$ evolve more or less independently from one
another in the avoided level crossing scenario \cite{rstopol} 
and that long-range correlations occur between the different states 
of the system \cite{jung}. 

The effective Hamiltonian used in
the random matrix interpolation between the standard
ensembles with real and complex matrix elements, is 
$H(\alpha) = H_0+\alpha H_1$ where $H_0$ and $H_1$ are real and complex random
hermitian matrices, respectively, and $\alpha$ is a crossover parameter
\cite{pandey}. For such an ensemble, the eigenvector elements acquire
correlations between the elements of the same eigenvector
\cite{efetov,brouwer1}  and between different eigenvectors 
\cite{brouwer4}. For individual systems, such a crossover may be observed
already in a billiard with only two attached waveguides \cite{sadreev2}.
The wave functions in the cross-over regime show long-range
correlations  \cite{brouwer3} like the eigenvectors of the Hamiltonian
$H(\alpha)$. These long-range correlations of the wave
functions in the real-to-complex crossover have recently been measured 
in an open microwave billiard \cite{brouwer5}.     

Thus, the results obtained experimentally and theoretically  for the profile
of the transmission peaks,  for the intensity fluctuations
and also for the long-range correlations of the wave functions
point to the  fact that the system through
which the transmission occurs may be essentially different from the original 
closed quantum system without attached leads. 
The differences arise from the interaction of neighboring resonance states 
via the continuum of scattering states. In quantum dots, these
differences can be traced experimentally
by varying the  gate voltage that controls the transparencies of the
point contacts. 

In the following, we will study the phases of the eigenfunctions of the
effective Hamiltonian  in the avoided level crossing scenario
in detail. We will show that they are not fixed but vary due to 
the mutial influence of neighboring 
resonance states.  In this spirit, a neighboring resonance state 
causes some "perturbation" for the  considered state
which is  similar to that caused by 
an impurity. There is however an important difference 
between these two cases. In contrast to the perturbation by an impurity,
the mutial interaction between
neighboring states can not be avoided. It is an internal property of an open
quantum system in the regime of overlapping resonances. 

In Sect. II, we repeat some general properties of the interplay between
the internal and external interaction in an open quantum system. While the
internal interaction is of standard type, the external interaction 
occurs additionally via the common continuum of scattering wave functions.  
The amplitude for the transmission through a quantum  dot will also be
given. It is resonant in relation to the effective Hamiltonian $H_{\rm eff}$
of the open quantum system. 
Then we define the phase rigidity $r_\lambda$ of the eigenfunctions 
$\phi_\lambda$ of $H_{\rm eff}$. Further, we consider  the scattering
wave functions $\Psi^E_C$ being solutions of $(H-E)\Psi^E_C=0$
in the total function space. In the interior of the system, the 
$\Psi^E_C$ can be represented by the $\phi_\lambda$ and, as a consequence,
the phase rigidity $\rho$ of the scattering wave functions $\Psi^E_C$
is related to that of  the eigenfunctions $\phi_\lambda$. 
In Sect. III, we provide numerical results for the relation between
transmission and phase rigidity $\rho$ of the scattering wave functions
for three special cases. The results are summarized and 
some conclusions are drawn in the last section.

\section{Phase rigidity of the eigenfunctions and of the scattering wave
function  in an open quantum system}

The relation between a closed and the corresponding open 
quantum system is as follows. 
A closed quantum system is described by a hermitian Hamilton operator $H_B$
the eigenvalues and eigenfunctions of which contain the  internal 
interaction $u$ of the discrete states. 
When embedded into the common continuum of scattering states, the
discrete eigenstates of the closed system 
turn over into resonance states with a finite lifetime.
The effective Hamiltonian $H_{\rm eff}$ of the open 
quantum system contains $H_B$ as well as an additional term \cite{rep}
that describes the coupling of the resonance states to the common environment,
\begin{eqnarray}
H_{\rm eff} = H_B + \sum_C V_{BC} (E^+ - H_C)^{-1} V_{CB} \; . 
\label{ham}
\end{eqnarray}
Here  $V_{BC}, ~V_{CB}$ stand for the coupling matrix elements between the 
{\it eigenstates} of $H_B$ and the environment that may consist of 
different continua $C$, e.g. the scattering waves propagating in the left 
($C=L$) and right ($C=R$) leads attached to a quantum dot \cite{saro}.
They are described by the Hamiltonian $H_C$. 
$H_{\rm eff}$ is non-hermitian,
its eigenvalues $z_\lambda$ and eigenfunctions $\phi_\lambda$ contain the
external interaction $v$ of the resonance states via the continuum
($v$ is used here instead of the concrete values  $V_{BC}$ and $V_{CB}$).
While $u$ causes level repulsion in energy, 
$v$ is responsible for the bifurcation of the widths of the resonance states. 

As long as $u/v \gg 1$, the spectroscopic properties of the open system are
similar to those of the corresponding closed system. The 
lifetimes (widths) of the resonance states are, 
as a rule, comparable in value, i.e. the states exist at the same time and 
can influence one another. 
Usually, the states avoid crossing
in the complex energy plane in a similar manner as it is well known 
from the avoided  crossings  of the discrete states of a closed system.  
When however $u/v \ll 1$, 
the resonance states do no longer exist at the same time due to their 
different lifetimes owing to widths bifurcation 
\cite{rep,rstopol}. They do not cross therefore in the complex energy plane.  
Most interesting is the situation  $u \approx v$. The interplay
between $u$ and $v$ may cause unexpected and even counterintuitive 
results such as the non-monotonic dependence of the resonance widths 
on the degree of opening the system 
\cite{marost,rstrans,rep,persson,perssonexp}. 

Since the effective Hamiltonian 
$H_{\rm eff}$ depends explicitely  on 
the energy $E$, so do its eigenvalues $z_\lambda$. The energy dependence is
weak, as a rule, in an energy interval that is determined by the 
width of the resonance state. 
The solutions of the fixed-point equations
$E_\lambda={\rm Re}(z_\lambda)_{|E=E_\lambda}$  
and of $\Gamma_\lambda=-2\, {\rm Im}(z_\lambda)_{|E=E_\lambda} $
are numbers that coincide approximately with the poles of the $S$ matrix. 
The width $\Gamma_\lambda$ determines
the time scale characteristic of the resonance state $\lambda$.
The amplitude for the transmission through a quantum dot is \cite{saro}
\begin{equation}
t=-2\pi
i\sum_{\lambda}\langle \xi^E_L|V|\phi_\lambda)
\frac{(\phi_\lambda|V|\xi^E_R\rangle }{E-z_{\lambda}}  
\label{trHeff}
\end{equation}
where the scattering wave functions in the leads are denoted by $\xi^E_C$.
According to (\ref{trHeff}), the transmission is {\it resonant} in relation 
to $H_{\rm eff}$. 
In (\ref{trHeff}), the eigenvalues $z_\lambda$ with their full 
energy dependence are involved.

The eigenfunctions $\phi_\lambda$  of $H_{\rm eff}$
are complex and  biorthogonal,
\begin{eqnarray} 
(\phi_\lambda|\phi_{\lambda '}) & \equiv & 
\langle\phi_\lambda^*|\phi_{\lambda '}
\rangle  =  \delta_{\lambda, \lambda '} 
\label{biorth1} \\
\langle\phi_\lambda|\phi_{\lambda}\rangle  \equiv  A_\lambda \ge 1
& ; & 
|\langle\phi_\lambda|\phi_{\lambda '}\rangle| \equiv B_\lambda^{\lambda '}
\ge 0 \; .
\label{biorth2}
\end{eqnarray}
The value
\begin{eqnarray} 
r_{\lambda}&=&\frac{\int dr ~\phi_{\lambda}^2}{\int dr
 |\phi_{\lambda}|^2} =  \frac{1}{ A_\lambda}
\nonumber \\
& =& 
\frac{\int dr (|{\rm Re}\, \phi_\lambda (r)|^2 - 
|{\rm Im}\, \phi_\lambda (r)|^2)}
{\int dr (|{\rm Re}\, \phi_\lambda (r)|^2 + |{\rm Im}\, \phi_\lambda (r)|^2)}
\label{rigbio}
\end{eqnarray}
is a measure for the biorthogonality of the eigenfunctions $\phi_\lambda$
of $H_{\rm eff}$ and for their phase rigidity, ~$0 \le r_\lambda \le 1$. It 
is large for an isolated resonance state at the energy  $E= E_\lambda$,
where the transmission probability has a peak. In the regime of overlapping 
resonance states however, where avoided level crossings appear \cite{rstopol},
$r_\lambda$ is usually small.
Approaching a branch point in the complex energy plane  where two 
eigenvalues $z_{\lambda}$ and $z_{\lambda '}$ coalesce,  we have
\cite{ro01,marost,rstopol} $A_{\lambda (\lambda ')} \to \infty$ and
$~r_{\lambda (\lambda ')} \to 0$, and further
\begin{eqnarray}
|\phi_{\lambda}) \to  \pm \; i\; |\phi_{\lambda '\ne \lambda}) \; .
\label{nl5}
\end{eqnarray}
Here, the widths bifurcate: with further increasing $v$,
one of the states aligns with a scattering state (channel wave 
function $\xi^E_{C=R}$ or  $\xi^E_{C=L}$ in the one-channel transmission) 
and becomes short-lived while the other one becomes 
long-lived. For large $v$, we have therefore only one long-lived resonance
state and,  as for non-overlapping resonance states, 
$r_\lambda \to 1$  at the energy  $E=E_\lambda $ of the long-lived state.

The scattering wave function $\Psi_C^E$ 
is solution of the Schr\"odinger equation 
$(H-E)\Psi^E_C=0 $ 
in the total function space that consists of the discrete states of the 
quantum dot and of the scattering states $\xi^E_C$ in the attached leads.
$H$ is hermitian. The solution reads \cite{rep}
\begin{equation}
\label{total}
\Psi_C^E(r)= \xi^E_C(r) + \sum_{\lambda} 
\Omega^C_\lambda (r) \,
\frac{(\phi_\lambda |V| \xi^E_C\rangle}{E-z_\lambda} 
\end{equation}
where  $\Omega^C_\lambda (r) =
\big[ 1+(E^{+}-H_C)^{-1} V\big]\phi_\lambda (r) $
is the wave function of the resonance state $\lambda$. 
According to (\ref{total}), the eigenfunctions $\phi_\lambda(r)$ of the 
effective Hamiltonian $H_{\rm eff}$ give the main
contribution to the scattering wave function $\Psi^E_C(r)$ in the interior 
of the cavity. Here,
\begin{eqnarray}
\Psi_C^E(r)\to \sum_\lambda \phi_\lambda (r) \,
\frac{(\phi_\lambda |V| \xi^E_C\rangle}{E-z_\lambda} \; .
\label{total1}
\end{eqnarray}

According to the relation (\ref{total1}), the phase rigidity $\rho$
of the scattering wave function,
Eq. (\ref{rig1}), is generally determined to the  values 
of the phase rigidity $r_\lambda$ of the individual resonance states 
$\lambda$. While $r_\lambda$  characterizes the phase rigidity 
of the special resonance state $\lambda$,  $\rho$ contains the 
$r_\lambda$ from the different states that contribute
to the scattering wave function 
$\Psi^E_C$ in the considered energy region. 
This difference between $\rho$ and $r_\lambda$
is illustrated best by the following example.  
Approaching a branch point in the complex energy plane, 
the contributions to (\ref{total1}) from
the two states $\lambda$ and $\lambda '$ with 
the relation (\ref{nl5}) of their wave functions to each other,
cancel each other. The phase rigidity $\rho$ of the scattering wave function 
is determined
therefore by the contributions of other states that, as a rule, 
are relatively far from one another. That means, in spite of 
$r_\lambda = r_{\lambda '} = 0$, $|\rho|$ might be large at the branch point.
 
We mention that the scattering wave function (\ref{total1}) has a similar
structure as the transmission amplitude (\ref{trHeff}). 
Both expressions consist of a sum over overlapping resonance states 
with the weight factor $(\phi_\lambda |V| \xi^E_C\rangle / (E-z_\lambda)$.
We expect therefore some correlation between both values.
This relation, being trivial for
isolated resonances, is of special interest in the regime of overlapping 
resonances. The transmission may be large not only at the positions 
$E=E_\lambda$  of the resonance states $\lambda$ as in the case of isolated
resonances. It may be enhanced in a larger energy region
when $|\rho| < 1$, i.e. when some states $\lambda$  are (partly) aligned
with the scattering states $\xi^E_C$ so that $r_\lambda <1$
for them. In such a case,    
the matrix elements $\langle \xi^E_C|V|\phi_\lambda)$ may be large
not only at $E=E_\lambda$ but in a larger energy region.  
For illustration, we will provide in the following section
some numerical results for $|t|$ and $|\rho|$ 
for two model quantum billiards with a small number of states 
the transmission through which 
is studied earlier, and for a realistic chaotic quantum billiard.  
The correlation between $|t|$ and $1-|\rho|$ 
in the overlapping regime can clearly be seen in all cases.

\section{Numerical examples}
\subsection{1d-system}

\begin{figure}[ht]
\includegraphics[width=4cm,height=3cm]{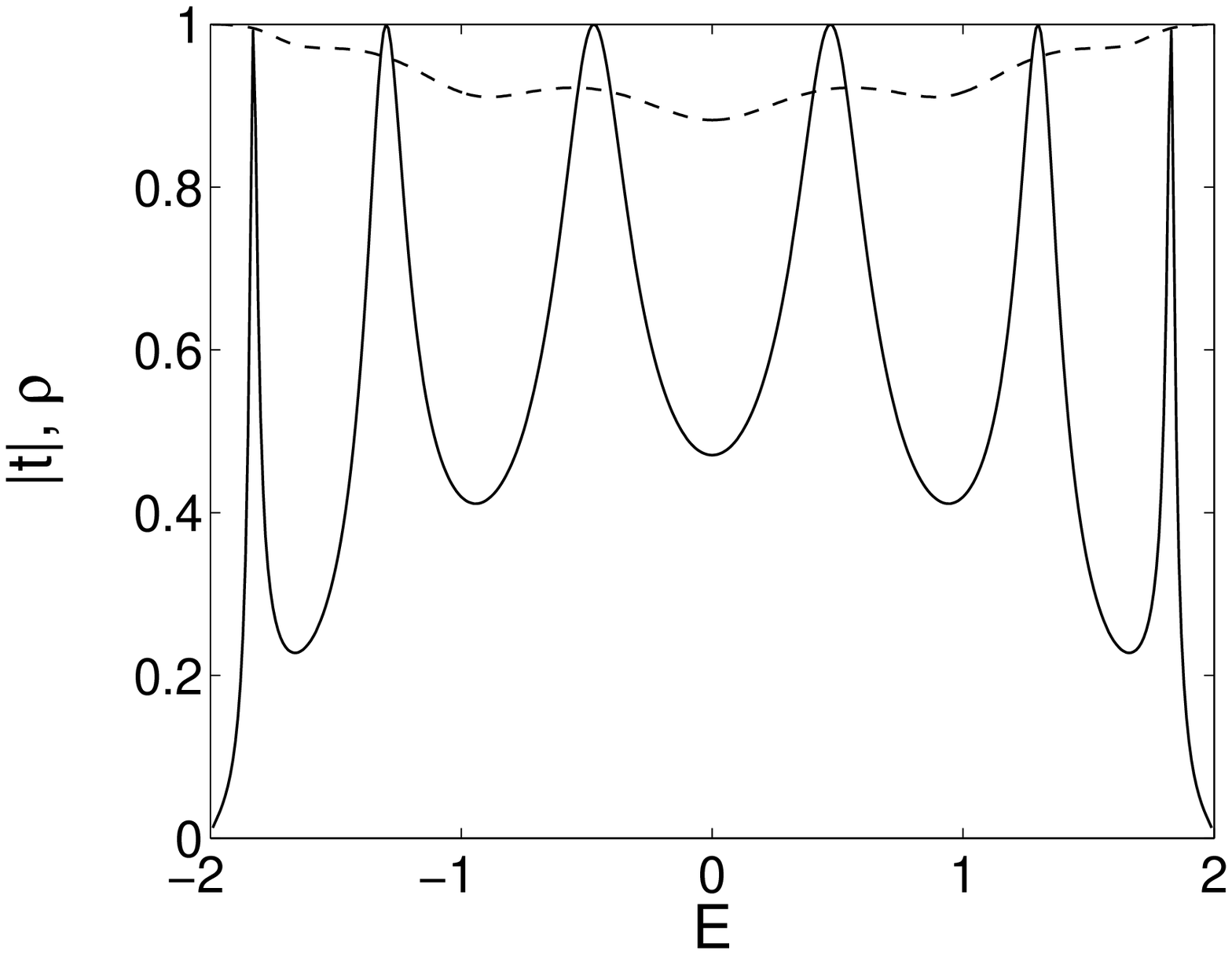} \hspace*{.3cm}
\includegraphics[width=4cm,height=3cm]{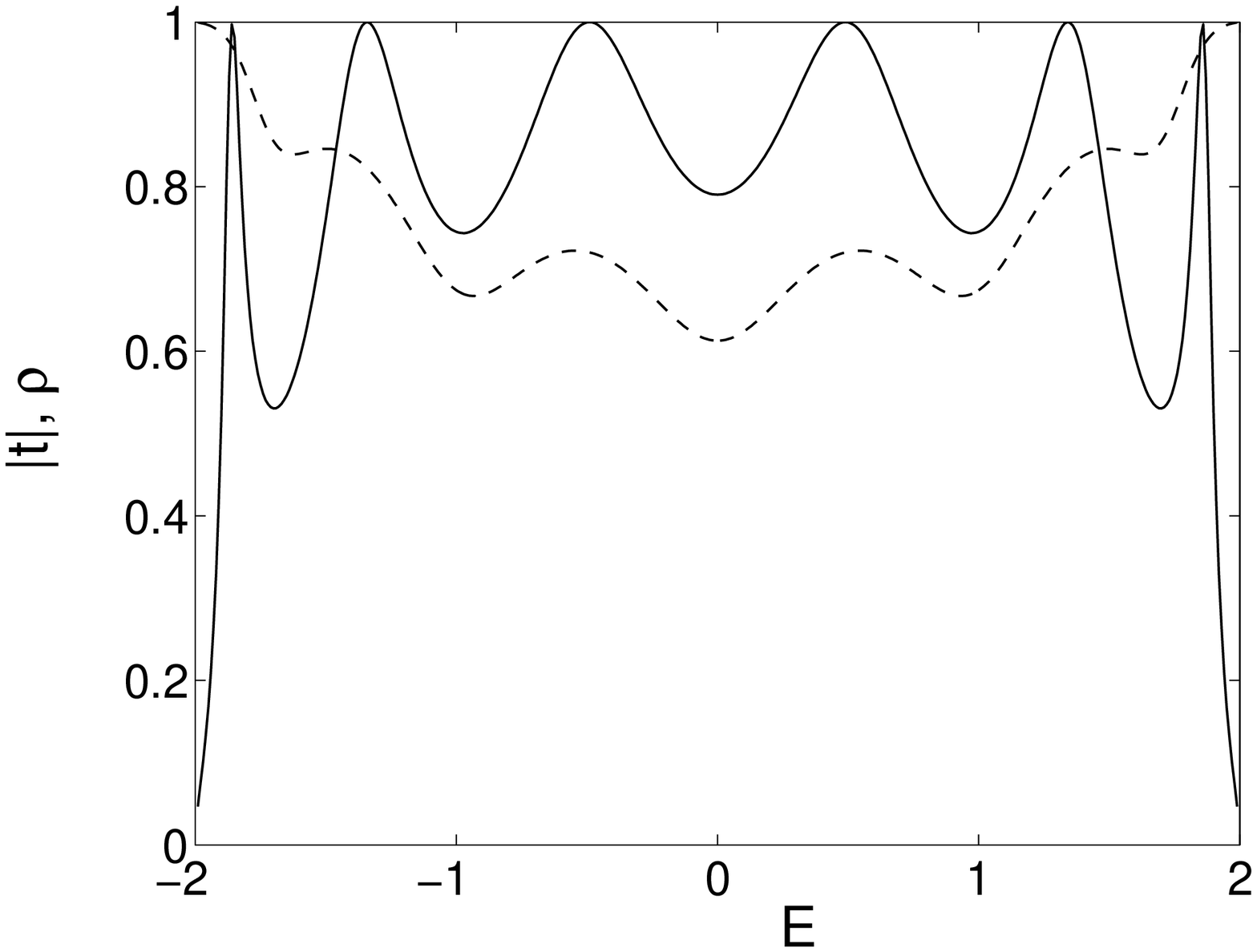} 
\caption{
The transmission  $|t|$ (full lines) and phase rigidity $|\rho|$ (dashed 
lines) for a chain of 6 sites  as a function of energy
with v=0.5 (left) and v=0.7 (right). For the model see Ref. \cite{saro}.
}
\label{fig1}
\end{figure}

For illustration, let us first consider the relation between the degree of 
resonance overlapping and the phase rigidity in the transmission
through a 1d-system, see Fig. \ref{fig1}. The widths of the 
states in the middle $E=0$ of the spectrum are larger than those near 
to the thresholds $E=\pm 2$ as can be seen immediately from
the profile of the transmission peaks. Therefore, also the degree of 
resonance overlapping increases towards the middle of the spectrum.
The results show that the transmission increases with increasing
resonance overlapping and, correspondingly, the  phase rigidity decreases.  
That means, the transmission increases when the resonance states align with
the scattering wave function which is expressed by the decreasing  
of $\rho$.

\subsection{Double quantum dot}

\begin{figure}[ht]
\includegraphics[width=6.2cm,height=5.6cm]{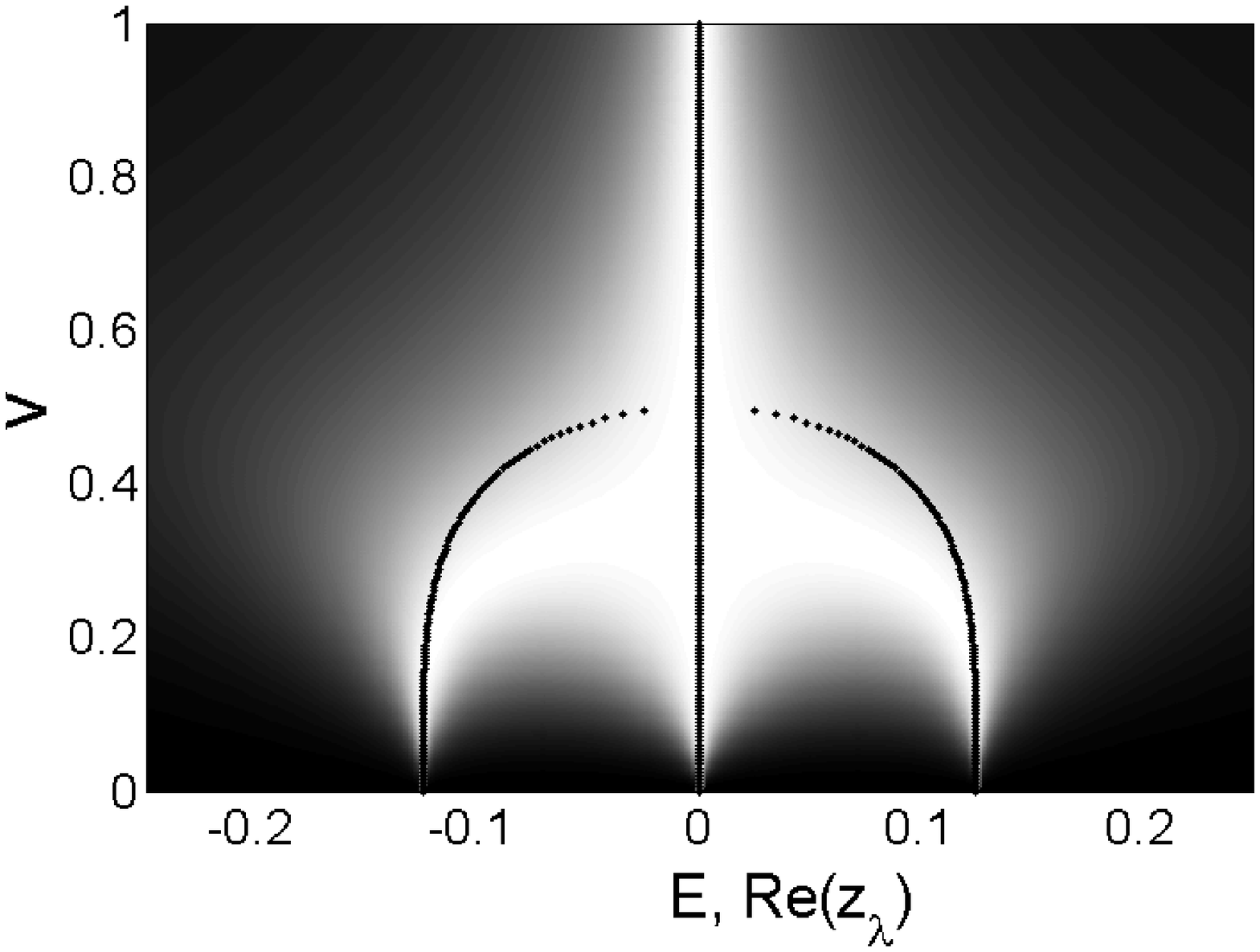} 
\includegraphics[width=6.2cm,height=5.6cm]{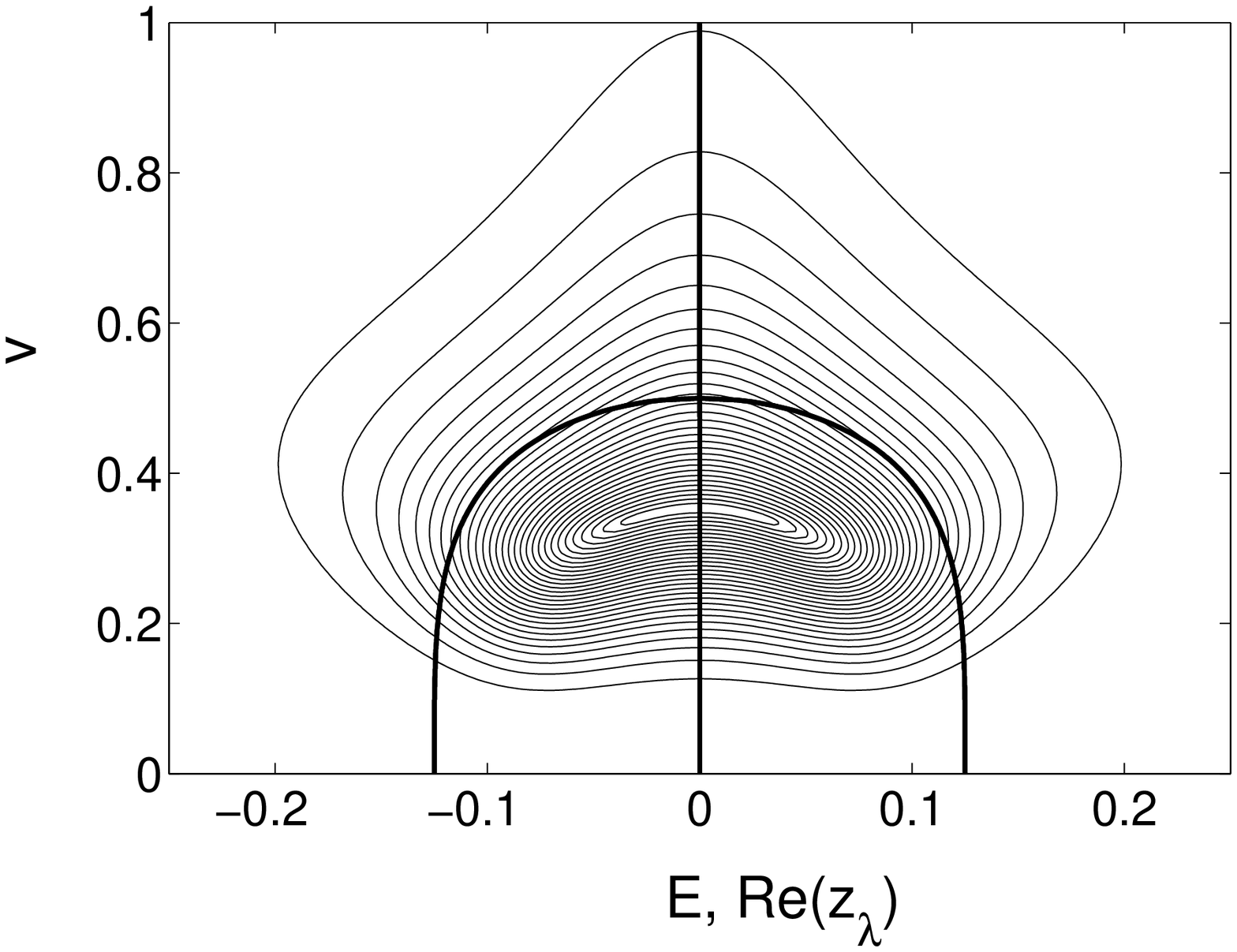} 
\caption{
The transmission $|t|$ (top) and the landscape of the phase
rigidity $|\rho|$ (bottom, thin lines) for a double quantum dot 
over energy $E$ and coupling strength $v$. 
The distance between the contour lines is $\Delta |\rho|
= 1/30 $. The minimal value $\rho = 0$ is surrounded by a high density of 
contour lines.   The highest shown contour line 
corresponds to $|\rho| = 1-1/30$. 
The  Re$(z_\lambda$) of the three eigenstates   (thick lines
in both panels of the figure) are calculated at $E=0$.
The branch point is at $v_c=1/2, ~E_c=0$. ~$u= \sqrt{2}/16 $. 
Around $v=0.345$, the phase rigidity is minimal and the transmission maximal 
with a plateau $|t|=1$ (for the plateau 
compare Fig. 4 in  Ref. \cite{rstrans}).
}
\label{fig2}
\end{figure}

Let us now consider a double quantum dot consisting of two
identical single dots that are connected by an internal wire. Suppose 
both dots and the wire have  one state each.
At $v=v_c$, the eigenvalues $z_\lambda$ and $z_{\lambda '}$
of the effective Hamiltonian  coalesce and a branch point
in the complex energy 
plane appears \cite{rstrans,rstopol}. When $v<v_c$,    the states 
repel each other in energy while widths bifurcation starts beyond the 
branch point where $v>v_c$. 
In Fig. \ref{fig2}, we show the transmission $|t|$ and the
landscape of $|\rho|$ over 
energy $E$ and coupling strength $v$ (for fixed $u$).
The smallest value of $|\rho|$ ($\rho =0$) is reached 
when the transmission is maximal with a plateau  $|t|=1$ (for the 
plateau compare Fig. 4 in  \cite{rstrans}). 
At $v=v_c$, ~$|\rho|$ is relatively large due to the fact that the 
contributions 
of the two states $\lambda$ and $\lambda '$ cancel each other. 
Beyond $v_c$, the phase rigidity increases further up to its maximal value 1.

\subsection{Sinai billiard}

\begin{figure}[ht]
\includegraphics[width=7.5cm,height=6.5cm]{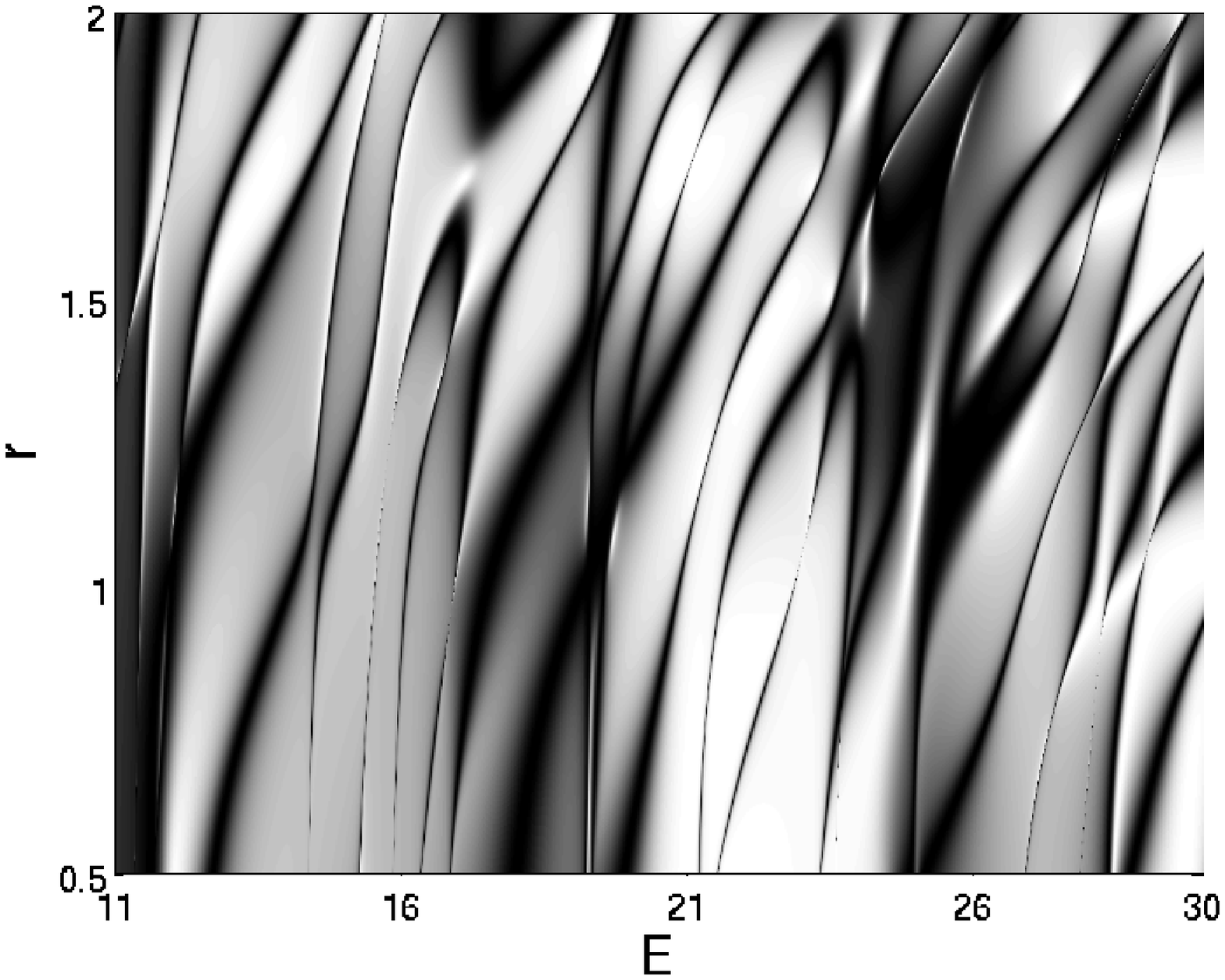} \\
\includegraphics[width=7.5cm,height=6.5cm]{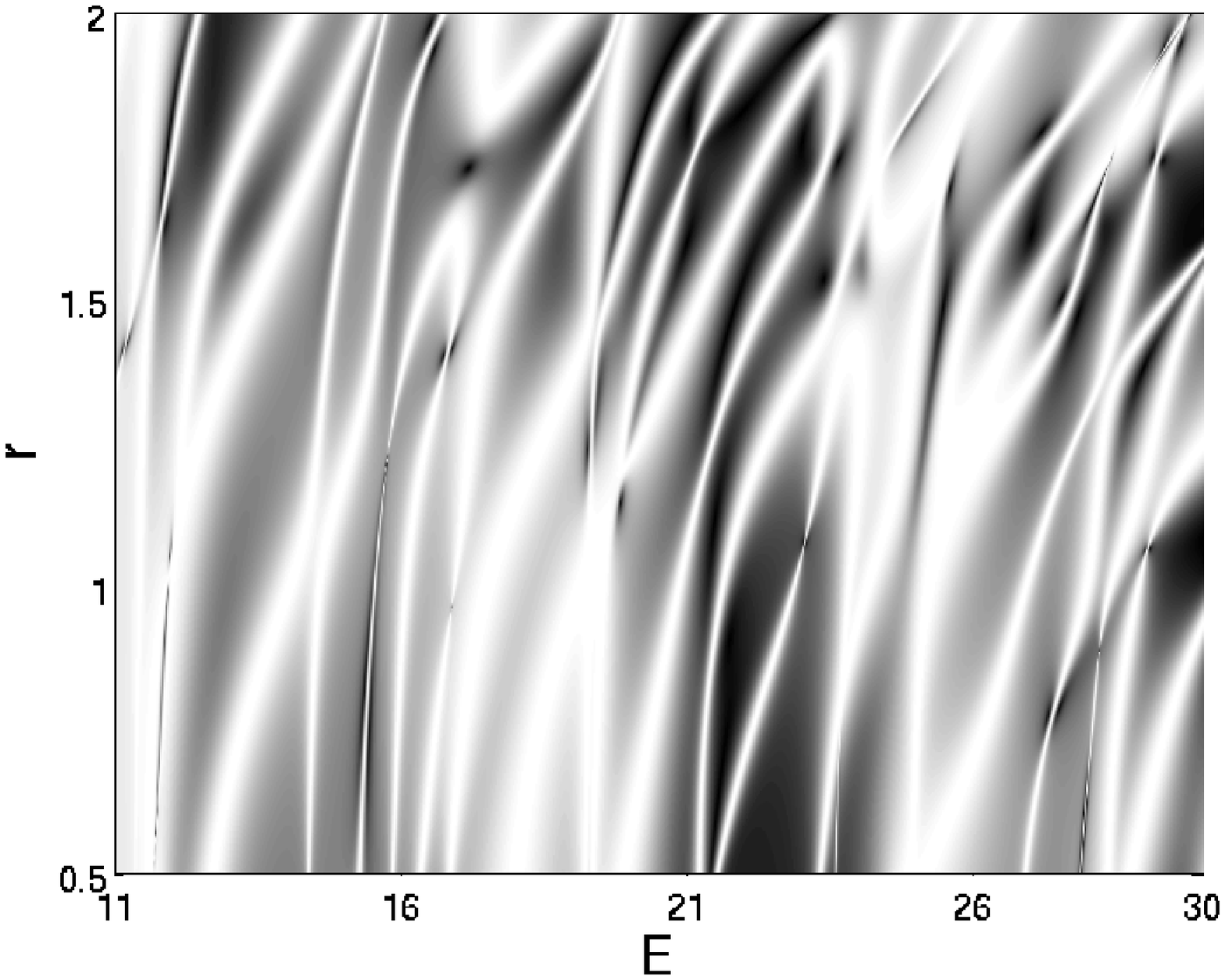} 
\caption{
The transmission $|t|$ (top) and phase rigidity $|\rho|$ 
(bottom) for a Sinai billiard over energy $E$ and radius $r$ of the 
circular disk (size of the billiard: $x= 4, ~y= 5$ in units of the 
width of the leads).
The calculations are performed in the tight-binding lattice model \cite{datta}.
}
\label{fig3}
\end{figure}

In Fig. \ref{fig3}, we show $|t|$ and $|\rho|$ calculated for a realistic 
chaotic system with many states and many avoided level crossings. 
The high correlation between the two values $|t|$ and $1-|\rho|$
can be seen also under these conditions. In some energy intervals,
$|t|$ is near to $1$. Here  $|\rho|$ is small, 
in agreement with the results shown in Fig. \ref{fig2}.

\section{Conclusions}

From the results obtained, we conclude 
that $|\rho| <1$ characterizes the scenario of overlapping resonances 
in which the resonance states interact with one another and 
avoided (and true) crossings appear, generally.
In this scenario, the system can not be described by a hermitian Hamilton
operator that provides rigid phases of its eigenfunctions ($r_\lambda =1$).  
Rather,  $H_{\rm eff}$ is non-hermitian, its eigenfunctions are biorthogonal
and $r_\lambda$ varies with energy and coupling strength.
Since the scattering wave function and
the eigenfunctions of the effective Hamiltonian are related to one another
according to  (\ref{total1}), also $\rho$ varies between 1
and 0 in the regime of overlapping resonances. That means, the reduction of
the phase rigidity $\rho$ of the scattering wave function is caused,
at least partly, by the mutial interaction of neighboring resonances. 
It is an internal property of an open quantum system in the overlapping 
regime and can not be avoided. 

The reduction of the phase rigidity $\rho$ is, in the one-channel transmission,
an expression for the (partial) alignment of two of the eigenfunctions 
$\phi_{\lambda,\lambda '}$ of the effective Hamiltonian $H_{\rm eff}$
with the two scattering wave functions $\xi^E_{C=R,L}$ 
in the overlapping regime. Due to this alignment (described by 
$r_{\lambda,\lambda '} <1$), the transmission through the system is enlarged
in a certain energy region around the energies $E_{\lambda,\lambda '}$ 
of the two resonance states. As can be seen from Fig.
\ref{fig2} and the corresponding discussion in \cite{rstrans}, 
the profile of the transmission is, when $\rho \approx 0$, 
completely different  from that
through two isolated resonance states. It is plateau-like. This result
explains the correlation between $|t|$ and $1-|\rho|$ which can be seen in all
our numerical studies (Figs. \ref{fig1}, \ref{fig2} and \ref{fig3}). 
Both values are, according to Eqs. (\ref{trHeff}) and (\ref{total1}),
characterized by the contributions from a sum of overlapping resonance states.

Since the variation of $r_\lambda$ and
$|\rho|$ with energy follows from the interaction of the
resonance states via the common continuum of scattering wave functions, 
it is related, generally,
to the profile of Fano resonances. The relation is however  
unique only with respect to $r_\lambda$.
For example, the mutial influence of neighboring resonances is,
according to the results presented in ~\cite{mudiisro,marost4},  
maximum when two eigenvalues of the effective Hamiltonian coincide.
The line profile of  two completely overlapping resonance states 
is, in the one-channel case  and up to the background term, 
given by \cite{ro03,marost4}
\begin{eqnarray}
S  = 1 - 2\, i \, \frac{\Gamma_d}{E-E_d + \, 
\frac{i}{2} \, \Gamma_d} -
\frac{\Gamma_d^2}{(E-E_d + \, \frac{i}{2} \, \Gamma_d)^2} 
\label{doublepole}
\end{eqnarray}
where $E_1 = E_2 \equiv E_d$ and $\Gamma_1 = \Gamma_2 \equiv \Gamma_d$.
According to this equation, the line profile of the two
resonances differs strongly from that of  isolated resonances  \cite{marost4}
(determined by the rigid value $r_\lambda =1$). 
In correspondence to this result, we have $r_\lambda =0$
in the one-channel case when two resonance states completely overlap. 
However  $|\rho|$ might be large 
in this case as for weakly overlapping resonance states, as discussed above. 
This result underlines once more the difference between the value $r_\lambda$ 
characteristic of a special resonance state $\lambda$ of the system,
and $\rho$ characteristic 
of the behavior of a sum of resonance states in the overlapping regime.

Thus, the  avoided crossings of resonance states represent one of the  
sources for the dephasing \cite{definition} observed in experimental data. 
The variation of $r_\lambda $ and $|\rho| $ between 0 and 1 is
characteristic of an open quantum system with overlapping 
resonance states, as stated above. Therefore, dephasing and dissipation 
should be equivalent in terms of their effect onto the Fano profile.
Such a result is observed experimentally, indeed \cite{stefan2}.

As to averaged values, we mention the following results obtained earlier.
The distributions of $|\rho|$ with energy and ensemble averaging  
calculated for large chaotic cavities, do not depend on
the concrete shape of the cavity \cite{mello} since the averaging   
smears the different contributions  to $|\rho|$.
As a result, the distribution of  $|\rho|$ is characterized, in such a case, 
only by the number of channels \cite{brouwer3}.

We underline that the interaction of neighboring
resonance states via the common continuum which is considered here
as source for the dephasing \cite{definition}, is fundamentally
different from, e.g., the electron-electron interaction or the interaction 
due to some radiation. It is rather of nonlinear geometrical origin, related 
to branch points  in the complex energy plane, as can be seen in the
following manner. On the one hand, the branch points 
determine the physical properties of an open quantum system in the overlapping
regime as discussed in \cite{ro01,rstrans,rstopol}. They are, generally, 
related to the avoided  crossings of resonance states which demonstrate
the mutual influence of resonance states.
On the other hand, as shown in the present paper, the 
mutual interaction of neighboring resonance states is
accompanied by  phase changes of the eigenfunctions
$\phi_\lambda$ of $H_{\rm eff}$ as well as of the scattering wave functions
$\Psi^E_C$ inside the cavity. In any case, the interaction of neighboring
resonance states via the common continuum and the resulting dephasing
can not be neglected in an open quantum system 
in the regime of overlapping resonances.

\vspace{.5cm}

{\bf Acknowledgments:} 
ENB and AFS  thank the MPI~PKS in Dresden for hospitality.
This work has been supported by RFBR grant 05-02-97713
"Enisey".

$^{*}$ 
ben@tnp.krasn.ru;
rotter$@$mpipks-dresden.mpg.de; \\
\hspace*{.5cm} almsa$@$ifm.liu.se;  ~almas$@$tnp.krasn.ru


\begin{thebibliography}{99}
\bibitem{definition} The term 'dephasing' is used here according to the 
definition given in Refs. \cite{brouwer6,brouwer2}.
\end{thebibliography}
\end{document}